# Decoupling of topology and texture in optical skyrmions under turbulence

D. G. Pires[*] and N. M. Litchinitser[†]

*Department of Electrical and Computer Engineering, Duke University, Durham, North Carolina 27708, USA*

*danilo.gomes.pires@duke.edu

[†]natalia.litchinitser@duke.edu

**Abstract**

Topological structure is widely invoked as a route to disorder-resilient photonic states, yet whether it protects locally resolved field structure under realistic disorder has not been established. Optical skyrmions, vectorial light fields characterized by a global skyrmion number $N_{sk}$, provide a stringent test of this question under turbulence. Although $N_{sk}$ is expected to be robust, conservation of a global invariant does not guarantee preservation of the underlying polarization texture. Here we reconstruct the full Stokes field of optical skyrmions transmitted through controlled turbulent channels, combining experiment, phase screen simulations, and analytical modelling to independently track global and local observables. We demonstrate a broad disorder regime in which $N_{sk}$ remains conserved while fine polarization structure rapidly degrades. This pronounced decoupling, strengthened for higher-order skyrmions, exposes a hierarchy of robustness between topological invariants and texture-resolved information, defining intrinsic limits of topological protection in disordered wave systems.

---

The ability to structure light in phase, polarization, and across space and time has opened new avenues for encoding information in its spatial and temporal degrees of freedom [1]. Many applications require propagation through environments that distort optical wavefronts, motivating the search for metrics that remain well-defined under perturbations [2]. Topology provides such metrics by treating an optical field as a mapping from physical coordinates to an internal state space that supports integer-valued invariants stable under smooth deformations [3]. These invariants offer compact labels of field structure that can persist even when conventional geometric measures fail. Atmospheric turbulence, however, introduces stochastic refractive-index fluctuations that cause scintillation, beam wander, and modal mixing [4–7], raising a central question: which structural metrics remain recoverable with sufficient fidelity after stochastic propagation?

Optical skyrmions provide a natural setting to address this question because they encode topology in polarization textures rather than scalar phase alone. Originally introduced in high-energy physics as topologically stable field configurations, skyrmions were later recognized as generic particle-like textures in ordered media [8–15]. In photonics, analogous structures can be engineered through superpositions of orthogonal polarization fields, periodic vector beams, spatiotemporal coupling, and fully three-dimensional vector fields [16–24]. This diversity makes them a versatile platform for testing the resilience of polarization-based topology relative to scalar structured beams [25–27]. Yet many applications rely not only on a conserved global invariant but also on the local texture that realizes it [18, 28, 29]. Here we disentangle these contributions using a controlled turbulent channel, statistically matched split-step simulations, and an analytical scaling model. We show that the global skyrmion index remains stable across a broad disorder regime while texture-resolved metrics degrade. This separation becomes increasingly pronounced with higher texture order, leading to reduced ensemble-averaged fidelity and enhanced variance. Our analytical model captures both the robustness of low-order textures and the rapid degradation of higher-order ones, establishing a hierarchy of stability within topological light.

**Global and local optical metrics distorted by turbulence**

Atmospheric turbulence arises from mixing air parcels with different temperatures and humidity, producing random refractive-index fluctuations along an optical propagation path [30, 31]. These inhomogeneities accumulate as stochastic optical-path variations that introduce phase delays and convert them into amplitude modulations, beam spreading, and intensity scintillation. A common statistical description is provided by Kolmogorov theory, which models the turbulence as locally homogeneous and isotropic over an inertial range between outer and inner length scales [32]. Turbulence strength is typically characterized by the refractive-index structure parameter $C_n^2$, enabling quantitative estimates of coherence loss and providing a basis for random phase-screen simulations. In the laboratory, turbulence is often emulated using programmable wavefront modulators driven by Kolmogorov statistics or with translated turbulence plates [33-36]. Here, we use a hot-air turbulence chamber to generate a distributed, dynamically evolving refractive-index distribution along the optical path [37]. Heater power and flow set the turbulence strength, which is calibrated from time-ensemble beam metrics to map experimental settings onto the Kolmogorov parameterization used in simulations (see Methods for chamber details).

These controlled realizations of atmospheric perturbations provide a direct route to investigate optical skyrmions, which are polarization textures defined by the spatial arrangement of the local polarization state across the beam cross-section. A convenient description maps each transverse point to a unit Stokes vector $\mathbf{s}(x,y) = \mathbf{S}/S_0$ on the Poincaré sphere, where $\mathbf{S} = (S_1, S_2, S_3)$ and $S_0$ are respectively the Stokes parameters and the total intensity. In this representation, the optical field forms a two-dimensional mapping into a closed order-parameter space. The associated skyrmion number is a global invariant that quantifies the degree of the mapping onto the sphere, and can be written as [16, 38]

$$N_{sk} = \frac{1}{4\pi} \iint_\sigma \mathbf{s} \cdot (\partial_x \mathbf{s} \times \partial_y \mathbf{s}) dx dy, \tag{1}$$

where the integrand is the local topological density and $\sigma$ is the integration region. This formulation reflects the global nature of the invariant, suggesting that the skyrmion number can remain stable under moderate distortions. However, preservation of the global index does not, by itself, ensure the stability of the detailed polarization texture that encodes practical information. Optical skyrmions can be realized by superposing two co-propagating modes in orthogonal circular polarizations, formed from a Gaussian and a Laguerre-Gaussian (LG) mode, respectively, as $\mathbf{E}(r, \phi) = A_0 \text{LG}_0^0(r, \phi) \hat{\mathbf{e}}_R + A_m \text{LG}_0^m(r, \phi) e^{i\delta} \hat{\mathbf{e}}_L$, where $\hat{\mathbf{e}}_{R,L}$ denote right- (RCP) and left-circular (LCP) polarization, $m$ is the azimuthal winding number, and $A_0, A_m, \delta$ control the relative normalized weights and phase, respectively (detailed in Methods section). Figure 1 (a) shows a schematic of the experimental setup used to generate and probe the skyrmionic textures in turbulence (see Methods for details). Examples of the measured unperturbed and perturbed optical skyrmions are shown in Fig. 1(b) for Néel (b1, b4), Bloch (b2, b5), and anti-skyrmion (b3, b6) textures, respectively.

Beyond the global index, skyrmion textures are also characterized by local metrics that quantify the spatial structure of the polarization state within the cross-section. Expressing the Stokes vector in terms of spherical angles $\beta, \gamma$ on the Poincaré sphere (see Section 1 of the Supplementary Materials for complete derivations) as $\boldsymbol{s} = (\sin\beta\cos\gamma, \sin\beta\sin\gamma, \cos\beta)$, one may define three local parameters [16, 38]. First, the polarity $p$, which characterizes the direction of the out-of-plane orientation near the texture center and at the boundary $\sigma$, is commonly defined as $P_{sk} = [\cos\beta(r_\sigma) - \cos\beta(0)]/2$. The second parameter is the vorticity $V_{sk} = (2\pi)^{-1}\int_0^{2\pi}\partial_\phi\gamma d\phi$, which quantifies the integer winding of the azimuthal angle $\gamma$ around a closed loop enclosing the core. Finally, the third parameter is the helicity $H_{sk}$, which specifies the orientation of the in-plane rotation of $(s_1, s_2)$ relative to the radial direction, and is defined as the offset between the texture angle and the geometric azimuth, $H_{sk} = \gamma - V_{sk}\phi$. Together, $(P_{sk}, V_{sk}, H_{sk})$ specify the local structure of the texture and distinguish different skyrmion types. For example, $(P_{sk}, V_{sk}, H_{sk}) = (1,1,0)$ and $(1,1,\pi)$ correspond to Néel-type skyrmions, $(1,1,\pm\pi/2)$ correspond to Bloch-type skyrmions, and $(1, -1, H_{sk})$ define anti-skyrmions for any $H_{sk}$.

Turbulence acts on these quantities through random refractive-index fluctuations that impose spatially varying phase delays and amplitude modulations, altering the relative phase and amplitude of the orthogonal polarization components across the beam. In physical terms, turbulence perturbs $\beta(x, y)$ and $\gamma(x, y)$ point by point, leading to core drift, local polarization rotation, and irregularities in the gradients $\partial_x \boldsymbol{s}$ and $\partial_y \boldsymbol{s}$. As a result, polarity becomes vulnerable when the area defining the core is displaced or obscured by intensity fluctuations. The vorticity is readily corrupted because it relies on coherent angular winding of $\gamma$, which turbulence can fragment by introducing spatially dependent relative-phase slips. The helicity is typically the most sensitive parameter, as it depends on a well-defined in-plane orientation that is strongly affected by local polarization distortions. On the other hand, the skyrmion number integrates the topological density over the full cross-section, so turbulence can strongly distort the local texture while leaving the net wrapping unchanged. Only sufficiently strong distortions produce large fluctuations in the integrated density or induce effective boundary leakage of the mapping. This separation between global robustness and local degradation motivates a focused examination of texture-level stability in turbulent channels.

**Robustness of skyrmion textures under turbulence**

Local texture metrics are extracted from the measured Stokes fields after turbulence by operating directly on $\boldsymbol{S} = (S_1, S_2, S_3)$ in the observation plane. The polarity $P_{sk}$ is determined from the circular component of the Stokes parameters, using the mean value of $S_3$ at the texture core and at a reference radius that samples the outer region of the pattern. In practice, we evaluate $S_3$ at $r = 0$ and at the boundary $r = r_\sigma$ and assign the polarity from the relative orientation of these values [16, 38]. The texture-border radius $r_\sigma$ was defined from the unperturbed skyrmion by selecting the radial location where the circular Stokes component $S_3$ undergoes its transition between $+1$ and $-1$, providing a consistent reference boundary for all turbulence realizations. This definition remains effective under moderate distortions, but it becomes ambiguous when turbulence displaces the apparent core or suppresses contrast near $r_\sigma$. To address this issue, we employ an ensemble over realizations and an intensity-tracking algorithm to recenter the distorted field distribution with respect to its weighted intensity centroid. Across all the experimentally and numerically obtained textures, 100 realizations were performed for each turbulence strength. For each realization, the intensity centroid $(x_m, y_m)$ is calculated upon evaluating $x_m = \iint xI(x,y)dxdy / I_0$ and $y_m = \iint yI(x,y)dxdy / I_0$, where $I(x,y)$ is the measured intensity and $I_0 = \iint I(x,y)\, dxdy$.

The vorticity $V_{sk}$ is obtained from the azimuthal winding of the transverse Stokes complex field $S_1$ and $S_2$, defined as $S_\perp = S_1 + iS_2 = |S_\perp|\exp(i\psi)$, where $\psi = \arg(S_\perp)$ is the in-plane polarization angle [16, 38]. We evaluate the integer circulation of $\psi$ around the texture center as $V_{sk} = (2\pi)^{-1} \int_0^{2\pi} \partial_\phi \psi d\phi$. This procedure is closely analogous to extracting the topological charge of an OAM mode from the circulation of the optical phase [39]. Turbulence introduces spatially varying phase perturbations that redistribute power among neighboring winding numbers, leading to crosstalk between adjacent modes [4, 5, 33, 36]. In the present case, the random perturbation of $\psi$ causes effective mixing among integer windings of the transverse polarization angle, so that the measured vorticity can fluctuate between realizations even when the global texture retains its overall topology. The analogy is therefore direct, where vorticity degradation represents the polarization-texture counterpart of OAM mode crosstalk, with the angle $\psi$ playing the role of the scalar optical phase.

The helicity $H_{sk}$ is estimated statistically by comparing the perturbed in-plane angle $\psi$ to the corresponding unperturbed reference $\psi_0$ at matched spatial locations. We define an angular difference $\Delta\psi = \text{wrap}(\psi - \psi_0)$ on $(-\pi, \pi]$ and model its distribution with a von Mises probability density [40], which is the circular analogue of a normal distribution, written as $f(\Delta\psi; \mu, \kappa) = (2\pi I_0(\kappa))^{-1} \exp[\kappa \cos(\Delta\psi - \mu)]$. Here, $\mu$ is the mean direction, $\kappa \geq 0$ is the concentration parameter, and $I_0$ is the zeroth-order modified Bessel function of the first kind. In this framework, the mean helicity is identified with the fitted $\mu$ (or, equivalently, with the argument of the first circular moment), while $\kappa$ quantifies how tightly the perturbed texture remains clustered around that mean. Specifically, if $\langle e^{i\Delta\psi} \rangle = Re^{i\mu}$, then $R \in [0,1]$ measures circular coherence and is mapped to $\kappa$ through the standard von Mises relation $R = I_1(\kappa)/I_0(\kappa)$. Turbulence broadens the $\Delta\psi$ distribution, which reduces $R$ and drives $\kappa$ downward, corresponding to a loss of a well-defined

$H_{sk}$ even when a nonzero mean $\mu$ can still be identified over the ensemble (see Section 3 of the Supplementary Materials for details on this statistical model).

Figure 2 summarizes the values of the skyrmion number $N_{sk}$, polarity $P_{sk}$, vorticity $V_{sk}$, and helicity $H_{sk}$ obtained numerically and experimentally for the first-order Néel- (a,d), Bloch- (b,e), and anti-skyrmion (c,f) textures over various turbulence strengths. The skyrmion number $N_{sk}$ was computed directly from Eq. (1). As the turbulence strength increases, the persistence of $N_{sk}$ alongside the rapid degradation of $(P_{sk}, V_{sk}, H_{sk})$ indicates that the polarization field retains its global wrapping on the Poincaré sphere even while the spatial organization that defines a recognizable texture is progressively disrupted. In other words, turbulence introduces transverse variations in the relative phase and amplitude between the orthogonal polarization components, which distort the gradients of the Stokes vector and redistribute the local topological density without necessarily altering its integral. This produces a regime where the global invariant remains measurable and close to its target value, yet the local structure required for encoding, classification, or tracking of textures becomes increasingly unreliable.

The loss of polarity at stronger turbulence has a direct geometrical meaning. The sign of $S_3$ at the core and at the edge $r_\sigma$ no longer provides a stable up/down orientation of the texture. Experimentally, this occurs when the core position becomes uncertain, when intensity fluctuations reduce the fidelity of the Stokes reconstruction near the origin, or when the radial profile is reshaped so that the reference ring no longer samples the same portion of the texture. Once polarity becomes ambiguous, the texture can no longer be assigned to a well-defined central orientation, undermining any interpretation that depends on identifying a core state or distinguishing textures that differ only by the sense of their out-of-plane component.

Vorticity degradation reflects a breakdown of a coherent azimuthal progression of the transverse Stokes-field phase, denoted by $\arg(S_1 + iS_2)$. With increasing turbulence, the winding number extracted around a closed contour becomes realization-dependent because phase perturbations induce angular discontinuities and create or displace points where $|S_\perp|$ is small, effectively fragmenting the circulation. The resulting histograms in Fig. 2(d–f) are therefore the polarization-texture analogue of mode mixing in OAM channels, where the intended winding remains statistically favored, but weight is transferred to neighboring integers as the random medium strengthens. This redistribution has practical consequences for any scheme that uses vorticity as an information carrier, since it directly translates into symbol errors, even when $N_{sk}$ still indicates that a nontrivial topology survives.

In addition, helicity exhibits a complementary form of fragility because it relies on a well-defined in-plane orientation relative to the reference texture. In the von Mises description, stronger turbulence broadens the distribution of angular differences, reducing the concentration parameter and driving the ensemble toward a nearly uniform angular distribution. In this regime, a mean helicity may still be formally defined, but it is no longer informative because the probability density is not localized around that mean value. This corresponds to a loss of a stable internal twist direction of the texture, such that the distinction between Néel-like and Bloch-like arrangements becomes progressively obscured, even though the global mapping still wraps the sphere approximately once.

These observations clearly separate topological robustness from texture fidelity. The resilience of $N_{sk}$ demonstrates that turbulence can preserve a global invariant even when polarity, vorticity, and helicity are no longer reliable metrics. Consequently, at stronger turbulence levels, the field may remain topologically nontrivial while carrying diminished usable texture information, suggesting that topology alone is insufficient to guarantee stability of the local features required for encoding and interpretation in realistic propagation channels.

**Topological degradation of higher-order optical skyrmions**

So far, measurements of first-order textures indicate that the global topological index can remain stable even after turbulence significantly distorts the spatial polarization pattern, supporting the use of $N_{sk}$ as a robust metric in disturbed channels. However, does this resilience extend to higher-order optical skyrmions? To address this question, we examined Néel-type textures with increasing target order, propagating them through the same turbulent channel and retrieving $N_{sk}$ in experiments and simulations. The combined results are shown in Fig. 3.

Across orders up to $N_{sk} = 5$, we find that the robustness of the skyrmion number is not universal. Instead, it becomes progressively reduced with increasing topological order. While low-order textures maintain near-constant $N_{sk}$ over turbulence levels that already disrupt the local texture metrics, higher-order textures exhibit a significantly faster loss of a well-defined skyrmion number. This is evidenced by broadened distributions and a reduced probability of recovering the target skyrmion number. This behavior is consistent with the fact that higher-order textures contain finer spatial variation of the Stokes field and a more structured topological-density distribution. As a result, turbulence more readily redistributes density across the aperture and amplifies cancellation between positive and negative contributions in the integral defined in Eq. 1. Consequently, the global invariant can remain informative for modest orders, but it becomes increasingly susceptible to turbulence as the texture complexity grows. This limits the range over which topological protection can be reliably leveraged. A comparable order-dependent reduction in skyrmion-number fidelity has been reported previously, with a greater spread for more complex textures at higher background noise levels [41].

To quantify the loss of skyrmion-number fidelity with increasing texture order, we evaluated the degradation through the ensemble average $\langle N_{sk} \rangle$ and variance $\text{Var}(N_{sk})$ over many turbulence realizations. These metrics capture the fluctuations of the retrieved integer around its target value and therefore provide a direct measure of reliability without being obscured by outliers. Across the explored orders, $\langle N_{sk} \rangle$ decreases, while $\text{Var}(N_{sk})$ increases rapidly with respect to its initially unperturbed skyrmion number $N_0$. This behavior reveals an exponential dependence that becomes evident already at moderate turbulence strengths, consistent with the observed broadening of the skyrmion-number distributions for higher-order textures. This trend is shown in Figs. 4 (a) and (b) for $\sigma_R^2 = 0.25$ across numerical simulations and experiments, respectively, corresponding to the ensemble average and variance of $N_{sk}$.

We developed a model, treating the effective cumulative distortions as a multiplicative modulation of the topological-density field prior to integration, to interpret this scaling. For the complete derivation, refer to Section 4 of the Supplementary Materials. Using the Rytov-Tatarski description of turbulence-induced field fluctuations [30, 31], the resulting estimate predicts that the ensemble average of the skyrmion number decays as $\langle N_{sk} \rangle = N_0 \exp(\Lambda_{eff} N_0^2)$ and its variance grows as $\text{Var}(N_{sk}) = N_0^2 \exp(\Gamma_{eff} N_0^2)$, where $\Lambda_{eff} < 0$ and $\Gamma_{eff} > 0$ are effective turbulence parameters. As seen in Fig. 4, this exponential dependence accurately describes both numerical and experimental measurements. It also formalizes the order-dependent limitation on skyrmion-number resilience, even when the mean value remains near the target value at low orders. The retrieved values from the simulations (experiments) were $\Lambda_{eff}^{sim} = -0.008668$ ($\Lambda_{eff}^{exp} = -0.008481$) and $\Gamma_{eff}^{sim} = 0.15$ ($\Gamma_{eff}^{exp} = 0.21$), respectively. Although this theoretical model was developed to understand the correlations between the skyrmion number and perturbations from atmospheric turbulence, it can be further modified to study other optical stochastic systems. Besides the clear analogy to the behavior of higher OAM scalar modes in turbulence [4, 5, 33, 36, 42], similar order-dependent limitations are reported in dilute Bose-Einstein condensates [43] and nematic liquid crystals [44]. In the present context, these precedents reinforce that increasing topological order does not automatically strengthen or maintain equal protection.

**Methods**

**Convective-air turbulence chamber**

The turbulence chamber used in the experiments generates a controlled volume of turbulent airflow with an internal mixing region measuring approximately $150 \times 150 \times 150$ mm$^3$ in a rectangular cuboid shape. Ambient air is driven into the chamber by two fans, while a third fan injects a heated stream; a separate exhaust fan maintains a steady throughflow and promotes continuous mixing. The optical beam enters and exits through dedicated apertures and experiences time-dependent wavefront distortions as refractive-index fluctuations evolve within the volume. The heated air source can reach 120°C with a temperature stability of approximately $\pm 0.1$°C. Under these operating conditions, the characteristic temporal bandwidth of the transverse flow corresponds to a Greenwood frequency of $f_G \approx 35$ Hz. It contains 4 circular optical ports with a diameter of about 76.2 mm located at each side of the chamber, which are used to reflect the structured light multiple times using mirrors.

**Optical skyrmionic textures**

Optical skyrmions were prepared as paraxial vector fields formed by the coherent addition of two spatial modes carried by orthogonal circular polarizations. The circular basis vectors are defined as $\hat{\mathbf{e}}_R = (\hat{\mathbf{x}} - i\hat{\mathbf{y}})/\sqrt{2}$ and $\hat{\mathbf{e}}_L = (\hat{\mathbf{x}} + i\hat{\mathbf{y}})/\sqrt{2}$. The skyrmionic complex envelope is written as [38, 41]

$$\mathbf{E}(r, \phi) = E_R(r, \phi, z)\,\hat{\mathbf{e}}_R + E_L(r, \phi, z)\,\hat{\mathbf{e}}_L, \tag{2}$$

with $E_R = A_0\,\mathrm{LG}_0^0(r, \phi, z)$ and $E_L = A_m\,\mathrm{LG}_0^m(r, \phi, z)\,e^{i\delta}$. Here, $\mathrm{LG}_0^0$ denotes the fundamental Gaussian mode, $\mathrm{LG}_0^m$ is an LG mode with azimuthal index $m$ and null radial orders, $A_0$ and $A_m$ set the relative normalized weights, and $\delta$ controls the relative phase between the two circular components. In standard notation, $\mathrm{LG}_0^m(r, \phi, z) = U_m(r, z)\,e^{im\phi}$, where $U_m(r, z)$ is the spatial envelope determined by the beam waist and propagation distance, given by [45]

$$U_m(r, z) = \sqrt{\frac{2}{\pi |m|!}} \frac{1}{w(z)} \left(\frac{\sqrt{2}r}{w(z)}\right)^{|m|} e^{-\frac{r^2}{w(z)^2}} e^{-i\frac{kr^2}{2R(z)}} e^{i(|m|+1)\zeta(z)}, \tag{3}$$

with the beam size $w(z) = w_0\sqrt{1 + (z/z_R)^2}$, wavefront radius of curvature $R(z) = z[1 + (z_R/z)^2]$, and the Gouy-phase-term factor $\zeta(z) = \operatorname{atan}(z/z_R)$. In all previous expressions, $z_R = kw_0^2/2$ is the Rayleigh range and $w_0$ is the minimum beam waist.

The resulting spatially varying superposition converts the azimuthal phase factor $e^{im\phi}$ into a polarization texture through the interference between the RCP and LCP components. From $\mathbf{E}$, the Stokes parameters $\mathbf{S} = (S_1, S_2, S_3)$ are obtained as $S_0 = |E_R|^2 + |E_L|^2$, $S_1 = |E_x|^2 - |E_y|^2$, $S_2 = |E_a|^2 - |E_d|^2$, and $S_3 = |E_R|^2 - |E_L|^2$. Here, $E_x, E_y, E_a, E_d$ correspond respectively to the horizontal, vertical, antidiagonal, and diagonal components of the vector field. The unit Stokes vector $\mathbf{s} = \mathbf{S}/S_0$ provides a map $(x, y) \to \mathbf{s}(x, y)$ that defines the skyrmion texture. The order of the texture is controlled primarily by the integer $m$, while the parameters $A_m/A_0$ and $\delta$ are chosen to set the radial extent of the transition in $S_3$ and the in-plane orientation of the transverse Stokes component, enabling reproducible preparation of the target skyrmion type at the input plane.

**Atmospheric phase screens**

Atmospheric turbulence was modeled numerically by sampling two-dimensional stochastic phase masks whose second-order statistics follow Kolmogorov scaling [30-32, 42, 46]. We define the phase structure function as $\Psi(\mathbf{r}) = \langle |\Phi(\mathbf{r}' - \mathbf{r}) - \Phi(\mathbf{r}')|^2 \rangle$, with $\langle \cdot \rangle$ denoting the ensemble average of a random phase screen $\Phi$, and relate it to a spatial-frequency power spectrum $\Theta(\boldsymbol{\kappa})$ through the standard Fourier relationship for stationary random processes. For Kolmogorov turbulence, the spectrum is evaluated from the Wiener spectrum $\Theta$ as [37, 42, 46]

$$\Psi(\mathbf{r}) = 2\int \Theta(\boldsymbol{\kappa})[1 - \cos(2\pi\boldsymbol{\kappa} \cdot \mathbf{r})]d^2\boldsymbol{\kappa} = 6.88\left(\frac{|\mathbf{r}|}{r_0}\right)^{5/3}, \tag{4}$$

where $\Theta(\boldsymbol{\kappa}) = 0.023 r_0^{-5/3}|\boldsymbol{\kappa}|^{-11/3}$, $r_0$ is the Fried parameter that sets the turbulence strength. Alternatively, $\Theta$ represents the covariance function of the random power spectrum $P(\boldsymbol{\kappa}) = \int \Theta(\boldsymbol{\kappa})\exp[-i2\pi\boldsymbol{\kappa}\cdot\mathbf{r}]d^2\mathbf{r}$. Phase screens are synthesized by a spectral sampling procedure on an

$N \times N$ grid spanning an aperture size $D$. A complex Gaussian random array $M(\boldsymbol{\kappa})$ (zero mean, unit variance) is generated on the discrete frequency lattice and shaped by the target spectrum as $P(\boldsymbol{\kappa}) = \sqrt{\Theta(\boldsymbol{\kappa})}\, M(\boldsymbol{\kappa})$ [46, 47]. The corresponding phase realization is obtained by an inverse Fourier transform, taking the real component, $\Phi(x, y) = \Re\{\mathcal{F}^{-1}[P(\boldsymbol{\kappa})]\}$. Because the smallest resolvable frequency on the grid is $1/D$, this direct sampling underrepresents large-scale contributions associated with the $\kappa^{-11/3}$ dependence. We therefore augment the screen with additional low-frequency components by evaluating the spectrum at frequencies below $1/D$ and adding the resulting long-period Fourier terms with independent random coefficients. This correction restores the slowly varying content of the phase maps and improves agreement with Kolmogorov statistics over the full range of spatial scales supported by the simulation grid.

**Experimental setup**

To quantify how the skyrmionic global and local optical metrics get affected by turbulence in experiments, a continuous-wave laser tuned at $\lambda = 532$ nm illuminates a spatial light modulator (Hamamatsu LCOS-SLM X15213-01) at the holographic modulation section, modulating the incoming beam into a pair of optical modes carrying a LG and Gaussian mode profiles, respectively. Here, a hologram is generated using an approximated inverse-sinc technique to achieve amplitude and phase control [48]. The generated modes are spatially superposed with orthogonal linear polarization upon inserting a half-wave plate (HWP) in one of the arms and filtered with an iris (I) to obtain a single LG mode in one arm and a single Gaussian mode in the other arm, then routed to propagate within the turbulence chamber. At the chamber entrance, a quarter-wave plate (QWP) converts the orthogonal linear polarization states into the corresponding circular polarization states, RCP and LCP. After the propagation, a beam splitter (BS) divides the beam into two routes, where a QWP is inserted in one arm. A polarization-resolved camera (Allied Vision Mako G-508B POL) is used to obtain the complete Stokes measurements at single shot, where the arm with the QWP allows the acquisition of the RCP/LCP states, while the other arm directly measures all four linear polarization components, *i.e.*, horizontal, vertical, diagonal, and anti-diagonal states, which are numerically superposed during the data analysis referenced by the intensity centroid, as discussed in the main text. This method of acquisition is particularly efficient for single-shot measurements of the Stokes parameters when time-varying media are used [49]. To emulate long propagation distances in turbulence, the Fresnel scaling number $F = w_0^2/\lambda L$ ensures that the experimental arrangement is compatible with the evolution of a $w_0 = 6$ mm beam size ($1/e^2$) over a $L = 270$ m turbulence channel [37, 42, 49]. Here, $w_0$ is the beam waist radius, $\lambda$ the wavelength of the light, and $L$ the channel length. In the laboratory frame, the distance of $L_{lab} = 1.5$ m is achieved by using mirrors to reflect the beam multiple times within the turbulence chamber at different internal positions.

**Numerical simulations**

Numerically, we reproduce the experiment using split-step propagation through multiple statistically independent phase screens drawn from a Kolmogorov spectrum, separated by free-space propagation segments. Here, the $w_0 = 6$ mm beam is propagated over the $L = 270$ m channel. The number of phase screens $N_{sc}$ is chosen so that each step corresponds to a weak interaction, while the cumulative distortion matches the target turbulence strength, following $N_{sc} = (10\sigma_R^2)^{-6/11}$ [50]. The Rytov variance is defined as $\sigma_R^2 = 1.23 C_n^2 k^{7/6} L^{11/6}$, where $k = 2\pi/\lambda$ is the wavenumber, while $C_n^2$ and $L$ were previously defined. To align laboratory settings with simulation parameters, the chamber is calibrated by monitoring intensity fluctuations of a Gaussian beam and estimating the values of $C_n^2$ from the beam wander at various temperature regimes. For more details, we present the calibration curve and the calculations used to estimate $C_n^2$ from experiments in Section 2 of the Supplementary Materials. The simulated fields are then processed with the same polarization reconstruction and topological analysis pipelines used for experimental measurements.

**Data availability.** The data that support the findings of this study are available from the corresponding author upon reasonable request.

**Code availability.** The code used to produce the figures is available from the corresponding author upon reasonable request.

**Acknowledgements.** This work was supported by the Army Research Office (W911NF2310057). We thank H. Barati Sedeh and J. Tan for the fruitful discussions.

**Author contributions.** N.L. and D.G.P. formulated the concept of this study, performed theoretical and numerical analysis, and interpreted the results. D.G.P. designed and operated the experiments. All authors contributed to the discussion of the results and writing of the manuscript.

**Competing interests.** The authors declare no competing interests.

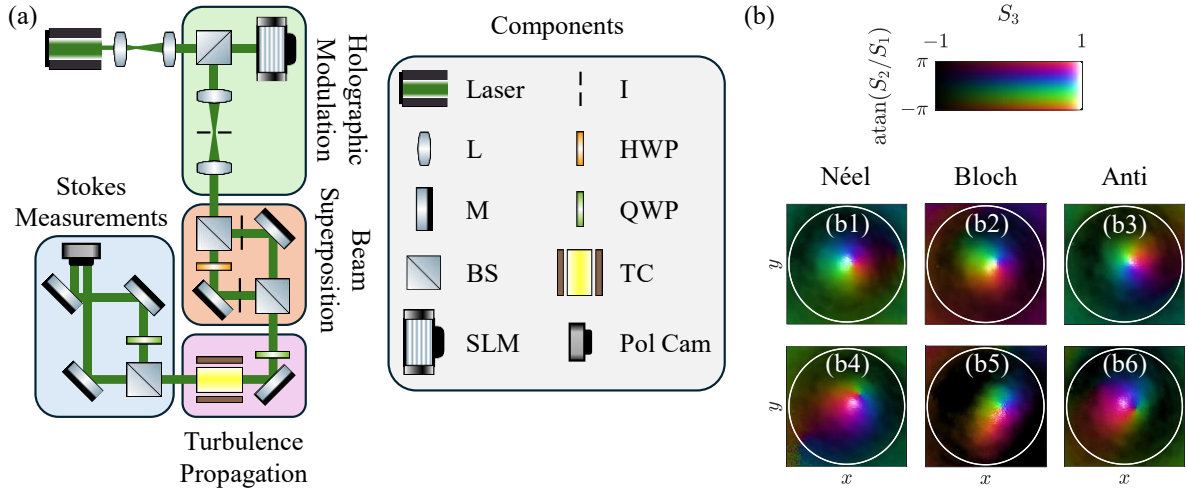

Fig 1. Experimental arrangement. a The LG and Gaussian modes are generated in experiments using a phase-only SLM in the Holographic Modulation section, overlapped in the Beam Superposition section, then routed to the chamber in the Turbulence Propagation section. Single-shot acquisition of the polarization information is realized in the Stokes Measurements section using a polarization-sensitive camera. The labels correspond to: L lens, M mirror, BS beamsplitter, SLM spatial light modulator, I iris, HWP half-wave plate, QWP quarter-wave plate, TC turbulence chamber, Pol Cam polarization-sensitive camera. b Examples of the measured Néel- (b1, b4), Bloch- (b2, b5), and anti-skyrmion (b3, b6) textures, where the first row indicates the distributions without perturbation and the second row corresponds to the cases for $\sigma_R^2 = 0.25$ ($T = 70°C$), respectively. The color scheme indicates the complete unit Stokes vector $\mathbf{s} = (s_1, s_2, s_3)$, and the white circle is the region of integration $\sigma$.

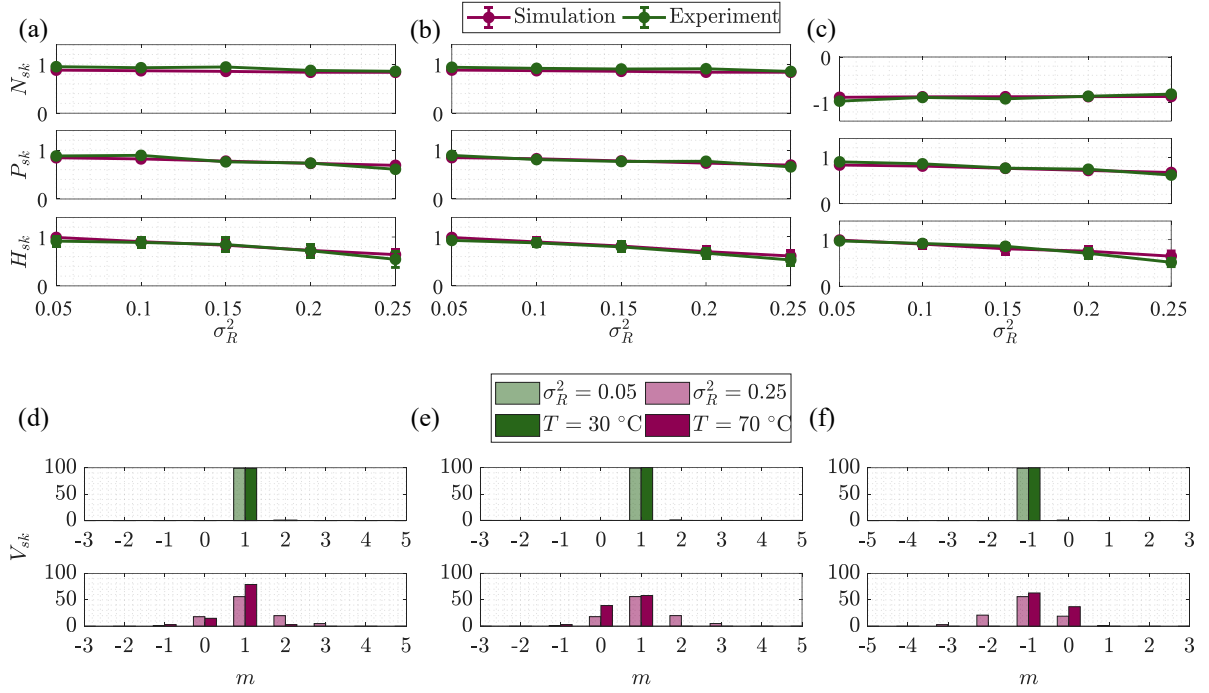

**Fig. 2. Robustness of global and local parameters to turbulence. a-c** Retrieved skyrmion number $N_{sk}$, polarity $P_{sk}$, and helicity $H_{sk}$ values from simulations (pink) and experiments (green) across various turbulence strength levels, denoted by the Rytov variance $\sigma_R^2$. From left to right, each panel corresponds to the Néel-, Bloch-, and anti-skyrmions, respectively. The error bars correspond to the standard error across the 100 realizations. **d-f** Histogram distributions for the vorticity $V_{sk}$ obtained for turbulence levels $\sigma_R^2 = 0.05$ ($T = 30°C$) and $\sigma_R^2 = 0.25$ ($T = 70°C$), here addressed by the green and pink colors, respectively. Light colors refer to simulations, while dark colors represent the experimental results. From left to right, each panel corresponds to the Néel-, Bloch-, and anti-skyrmions, respectively. In all histograms, the parameter $m$ denotes the measured winding number across the statistical realizations.

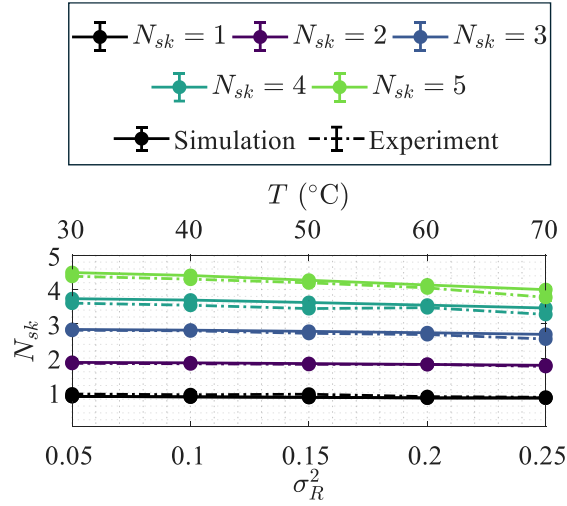

**Fig. 3. Stability of higher-order topologies.** Skyrmion number $N_{sk}$ inferred from simulations (solid lines) and experiments (dashed lines) for various topological orders, propagating a Néel-type skyrmion with skyrmion number spanning from $N_{sk} = 1, \ldots, 5$. The Rytov variance axis $\sigma_R^2$ corresponds to the turbulence strength parameter for the simulation results, while the temperature $T$ axis refers to the experiments. The error bar is the standard error across 100 realizations for each point.

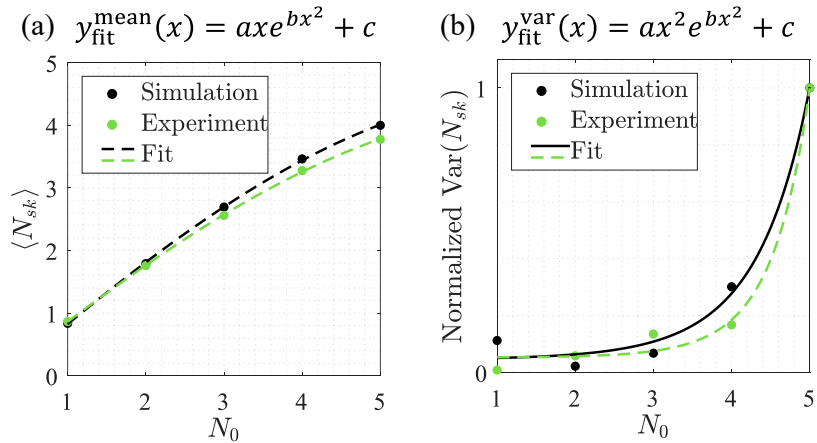

**Fig. 4. Uncertainty of higher-order skyrmion number predictions.** (a) Ensemble average and (b) normalized variance of the skyrmion number values ensembled across simulation (black dots) and experiment (green dots) for $N_0 = 1, \ldots, 5$ and turbulence level $\sigma_R^2 = 0.25$ ($T = 70°C$) from Fig. 3. The fitting curves for the simulation (black solid line) and experimental (dashed green line) results with function (a) $y(x) = axe^{bx^2} + c$ and (b) $y(x) = ax^2 e^{bx^2} + c$, respectively showing an exponential response in the variance of retrieved skyrmion numbers with respect to the initial topological order. In panel (b), the simulation and experimental results are normalized to their own maximum $\text{Var}(N_{sk})$ values.